\documentclass[letterpaper]{jpconf}
\usepackage{graphicx}
\usepackage{amssymb,amsmath,amsbsy}
\newenvironment{Eqnarray}%
     {\arraycolsep 0.14em\begin{eqnarray}}{\end{eqnarray}}

\newcommand\longdash{\hbox{\rm{\phantom{a}---\phantom{a}}}}
\def\beqa{\begin{Eqnarray}}
\def\eeqa{\end{Eqnarray}}
\def\beq{\begin{equation}}
\def\eeq{\end{equation}}
\def\hsm{h^0}
\def\mhsm{m_h}
\def\hh{H^0}
\def\hl{h^0}
\def\ha{A^0}

\def\mhh{m_{H}}
\def\mhl{m_{h}}
\def\mha{m_{A}}
\def\mhpm{m_{H^\pm}}
\def\mz{m_Z}
\def\mw{m_W}
\def\half{\tfrac{1}{2}}
\def\ifmath#1{\relax\ifmmode #1\else $#1$\fi}
\def\ls#1{\ifmath{_{\lower1.5pt\hbox{$\scriptstyle #1$}}}}
\def\Ref#1{ref.~\cite{#1}}
\def\lsim{\mathrel{\raise.3ex\hbox{$<$\kern-.75em\lower1ex\hbox{$\sim$}}}}
\def\gsim{\mathrel{\raise.3ex\hbox{$>$\kern-.75em\lower1ex\hbox{$\sim$}}}}
\def\phm{\phantom{-}}
\def\tanb{\tan\beta}

\begin{document}

\begin{flushright}
SCIPP-10/17\\
November, 2010\\[-1mm]
\end{flushright}
\vspace*{-0.7cm}

\title{Present status and future prospects for a Higgs boson
discovery at the Tevatron and LHC}

\author{Howard E. Haber}

\address{Santa Cruz Institute for Particle Physcs \\ 
University of California, Santa Cruz, CA 95064 USA 
}

\ead{haber@scipp.ucsc.edu}

\begin{abstract}
Discovering the Higgs boson is one of the primary goals of both the
Tevatron and the Large Hadron Collider (LHC).  The present status of
the Higgs search is reviewed and future prospects for discovery at
the Tevatron and LHC are considered.  This talk focuses primarily on
the Higgs boson of the Standard Model and its minimal supersymmetric
extension.  Theoretical expectations for the Higgs boson and its
phenomenological consequences are examined.

\end{abstract}

\section{Introduction}
\bigskip

The origin of electroweak symmetry breaking and the mechanism that
generates the masses of the known fundamental particles
is one of the central challenges of particle physics.
The Higgs mechanism~\cite{Higgs-orig} in its most general form
can be used to explain the observed masses of the $W^\pm$
and $Z$ bosons as a consequence
of three Goldstone bosons ($G^\pm$ and $G^0$) that
end up as the longitudinal components of the massive gauge bosons.
These Goldstone bosons are generated by
the underlying dynamics responsible for electroweak symmetry breaking.
However, the fundamental nature of this dynamics is still unknown.
Two broad classes of electroweak
symmetry breaking mechanisms have been pursued theoretically.  In one
class of theories, the electroweak symmetry breaking dynamics is
weakly-coupled, while in the second class of theories the dynamics is
strongly-coupled~\cite{weinberg}.

The electroweak symmetry
breaking dynamics of the Standard Model (SM) employs a
self-interacting complex doublet of scalar fields, which consists of four
real degrees of freedom~\cite{hhg}.  Elementary scalar dynamics 
yields a non-zero scalar field vacuum expectation value (vev), which leads to
the breaking of the electroweak symmetry.
Consequently, three massless Goldstone bosons are generated, while
the fourth scalar degree of freedom that remains in the physical spectrum
is the CP-even neutral Higgs boson of the Standard Model.
Precision electroweak data favors a Higgs mass below
200~GeV, in which case the scalar self-interactions are 
weak~\cite{LEPEWWG,GFITTER}.
In the weakly-coupled approach to electroweak symmetry breaking, the
Standard Model is very likely embedded in a supersymmetric
theory~\cite{susyreview} in
order to stabilize the large gap between the electroweak and the Planck
scales in a natural way~\cite{natural}.
The minimal supersymmetric extension of the Standard Model (MSSM)
employs two complex Higgs doublets, resulting in five physical scalar
degrees of freedom.  In a large range of the MSSM parameter space, the 
properties of the lightest scalar of the MSSM are nearly indistinguishable
from those of the SM Higgs boson.

It is possible that the dynamics responsible for electroweak symmetry
breaking is strongly-coupled~\cite{hill}.  Precision electroweak
data disfavors the simplest classes of models based on this approach.
Nevertheless, it remains possible that the physics of electroweak
symmetry breaking is more complicated than suggested by the Standard
Model or its supersymmetric extensions.  However, in this talk I will
not consider these theoretical alternatives nor their phenomenological
consequences.

\section{Theory of weakly-coupled Higgs bosons}
\bigskip
\subsection{The Standard Model Higgs Boson}
\medskip

In the Standard Model, the Higgs mass is given by: $\mhsm^2=\half\lambda
v^2$, where $\lambda$ is the Higgs self-coupling parameter.  Since
$\lambda$ is unknown at present, the value of the Standard Model Higgs
mass is not predicted.  In contrast, the Higgs couplings to
fermions [bosons] are predicted by the theory to be
proportional to the corresponding particle masses [squared-masses].
In particular, the SM Higgs boson is a CP-even scalar, and its
couplings to gauge bosons, Higgs bosons and
fermions are given by:
\beqa 
&& g_{h f\bar f}= \frac{m_f}{v}\,,\qquad\qquad
\qquad\qquad
g_{hVV} = \frac{2m_V^2}{v}\,, \qquad\qquad
g_{hhVV} = \frac{2m_V^2}{v^2}
\,,\label{hsmcouplings1}\nonumber \\[5pt]
&& g_{hhh} = \tfrac{3}{2}\lambda v =\frac{3\mhsm^2}{v}\,,\,\,\,
\quad\qquad
g_{hhhh} = \tfrac{3}{2}\lambda= \frac{3\mhsm^2}{v^2}\,,\label{hsmcouplings2}
\eeqa
where $V=W$ or $Z$ and $v=2m_W/g=246$~GeV.
In Higgs production and decay processes, the dominant mechanisms involve
the coupling of the Higgs boson to the $W^\pm$, $Z$ and/or
the third generation quarks and leptons.
Note that a $\hsm gg$ coupling ($g$=gluon)
is induced by virtue of a one-loop graph
in which the Higgs boson couples dominantly to a virtual $t\bar t$ pair.
Likewise, a $\hsm\gamma\gamma$ coupling is generated, although in this
case the one-loop graph in which the Higgs boson couples to
a virtual $W^+W^-$ pair is the dominant contribution.   Further details
of the SM Higgs boson properties are given in \Ref{hhg}.  
Reviews of the SM Higgs properties and its phenomenology, with an emphasis on
the impact of loop corrections to the Higgs decay rates and
cross-sections can be found in Refs.~\cite{kniehlreview,carenahaber,djouadi1}.

\subsection{Extended Higgs sectors}
\medskip

For an arbitrary Higgs sector, the tree-level $\rho$-parameter is
given by~\cite{tsao} 
\beq 
\rho\ls 0\equiv\frac{m_W^2}{m_Z^2\cos^2\theta_W}=1
\quad\Longleftrightarrow\quad (2T+1)^2-3Y^2=1\,, 
\eeq 
independently of the Higgs vevs, where $T$ and $Y$ specify the
weak-isospin and the hypercharge of the Higgs representation to which
it belongs.  $Y$ is normalized such that the electric charge of the
scalar field is $Q=T_3+Y/2$.  The simplest solutions are Higgs
singlets $(T,Y)=(0,0)$ and hypercharge-one complex Higgs doublets
$(T,Y)=(\half,1)$.  Thus, we shall consider non-minimal Higgs sectors
consisting of multiple Higgs doublets (and perhaps Higgs singlets),
but no higher Higgs representations, to avoid the fine-tuning of Higgs
vevs.

The two-Higgs-doublet model (2HDM) consists of two hypercharge-one scalar
doublets.  Of the eight initial degrees of freedom, three correspond
to the Goldstone bosons and five
are physical: a charged Higgs pair, $H^\pm$ and three neutral scalars.
In contrast to the SM where the Higgs-sector is CP-conserving, 
the most general 2HDM allows for
Higgs-mediated CP-violation (see \Ref{deva} for a complete 
specification of the 2HDM tree-level interactions). 
If CP is conserved, the Higgs spectrum
contains two CP-even scalars, $h^0$ and $H^0$
and a CP-odd scalar~$A^0$.
Thus, new features of the extended Higgs sector include:
(i) charged Higgs bosons; (ii) a
CP-odd Higgs boson (if CP is conserved in the Higgs sector); (iii) possible
Higgs-mediated CP-violation and neutral Higgs states of indefinite CP.
More exotic Higgs sectors allow for doubly-charged Higgs bosons, etc.
Further details 
of extended Higgs sectors can be found in \Ref{hhg}. 

\subsection{The Higgs sector of the MSSM}
\medskip

The Higgs sector of the MSSM is a 2HDM, whose Yukawa couplings and
Higgs potential are constrained by supersymmetry (SUSY).
Minimizing the Higgs potential, the neutral components of the
Higgs fields acquire vevs $v_u$ and $v_d$, where
$v^2\equiv v_d^2+v_u^2={4\mw^2/ g^2}=(246~{\rm GeV})^2$.
The ratio of the two vevs is an important parameter of the model,
$\tan\beta\equiv v_u/v_d$.
The five physical Higgs particles
consist of a charged Higgs pair $H^\pm$,
one CP-odd scalar $A^0$,
and two CP-even scalars $h^0$, $H^0$, obtained by diagonalizing
the $2\times 2$ CP-even Higgs
squared-mass matrix.
All tree-level Higgs masses and couplings and the 
angle $\alpha$ (which parameterizes the diagonalization
of the  $2\times 2$ CP-even Higgs
squared-mass matrix)
can be expressed in terms of two Higgs sector
parameters, usually chosen to be $\mha$ and $\tan\beta$.
See refs.~\cite{carenahaber,djouadi2} for further details.

At tree level,
$\mhl\leq\mz |\cos 2\beta|\leq\mz$,
which is ruled out by LEP data.  But, this
inequality receives quantum corrections.  The tree-level Higgs mass is shifted
due to an incomplete cancellation of 
particles and their superpartners (which would be an exact
cancellation if supersymmetry were unbroken)
that contribute at one-loop to the Higgs self-energy function.
The Higgs mass upper bound is then modified to~\cite{hradcorr}
\beq
\mhl^2\lsim \mz^2+{3g^2 m_t^4\over
8\pi^2\mw^2}\left[\ln\left({M_S^2\over m_t^2}\right)+{X_t^2\over M_S^2}
\left(1-{X_t^2\over 12M_S^2}\right)\right]\,,
\eeq
where $X_t\equiv A_t-\mu\cot\beta$ governs stop mixing
and $M_S^2$ is the average top-squark squared-mass.
The state-of-the-art computation includes the full one-loop result,
all the significant two-loop contributions, 
some of the leading three-loop terms, and renormalization-group
improvements~\cite{hradcorr2}.  
The final conclusion is that $\mhl\lsim 130$~GeV
[assuming that the top-squark mass is no heavier than about 2~TeV].
This upper bound is reached when $\tan\beta\gg 1$ and $\mha\gg m_Z$ 
in the so-called \textit{maximal mixing
scenario}, which corresponds to choosing $X_t/M_S\sim 2$ 
so that the predicted value of $\mhl$ is maximal.

Tree-level couplings of the MSSM
Higgs bosons with gauge bosons are often suppressed by
an angle factor, either $\cos(\beta-\alpha)$ or $\sin(\beta-\alpha)$,
as shown in the table below.
$$
\renewcommand{\arraycolsep}{1cm}
\let\us=\underline
\begin{array}{lll}
\us{\cos(\beta-\alpha)}&  \us{\sin(\beta-\alpha)} &
\us{\hbox{\rm{angle-independent}}} \\ [3pt]
\noalign{\vskip3pt}
       \hh W^+W^-&        \hl W^+W^- &  \qquad\longdash   \\
       \hh ZZ&            \hl ZZ  & \qquad\longdash  \\
       Z\ha\hl&          Z\ha\hh  & ZH^+H^-\,,\,\,\gamma H^+H^-\\
       W^\pm H^\mp\hl&  W^\pm H^\mp\hh & W^\pm H^\mp\ha \\
\end{array}
$$
Tree-level Higgs-fermion couplings may be either suppressed or enhanced with
respect to the SM value, $gm_f/2\mw$.  The charged Higgs boson couplings to fermion pairs,
with all particles pointing into the vertex, are:
\beqa
g_{H^- t\bar b}= &&~~~ \frac{g}{\sqrt{2}\mw}\
\Bigl[m_t\cot\beta\,P_R+m_b{\tan\beta}\,P_L\Bigr]\,,
\nonumber\\
g_{H^- \tau^+ \nu}= &&~~~\frac{g}{\sqrt{2}\mw}\
\Bigl[m_{\tau}{\tan\beta}\,P_L\Bigr]\,,\nonumber
\eeqa
where $P_{R,L}\equiv\half(1\pm\gamma_5)$,
and the neutral Higgs boson couplings are:
\beqa
\hl b\bar b \;\;\; ({\rm or}~ \hl \tau^+ \tau^-):&&~~ -
{\sin\alpha\over\cos\beta}=\sin(\beta-\alpha)
-\tan\beta\cos(\beta-\alpha)\,, \nonumber \\
\hl t\bar t:&&~~~ \phm{\cos\alpha\over\sin\beta}=\sin(\beta-\alpha)
+\cot\beta\cos(\beta-\alpha)\,, \nonumber \\
\hh b\bar b \;\;\; ({\rm or}~ \hh \tau^+ \tau^-):&&~~~
\phm{\cos\alpha\over\cos\beta}=
\cos(\beta-\alpha)
+\tan\beta\sin(\beta-\alpha)\,, \nonumber \\
\hh t\bar t:&&~~~ \phm{\sin\alpha\over\sin\beta}=\cos(\beta-\alpha)
-\cot\beta\sin(\beta-\alpha)\,, \nonumber \\
\ha b \bar b \;\;\; ({\rm or}~ \ha \tau^+
\tau^-):&&~~~\phm\gamma_5\,\tan\beta\,, \nonumber \\
\ha t \bar t:&&~~~\phm\gamma_5\,{\cot\beta}\,, \nonumber
\eeqa
where the $\gamma_5$ indicates a pseudoscalar coupling.
Especially noteworthy is the possible $\tan\beta$-enhancement
of certain Higgs-fermion couplings.  Typically,
 $1\lsim\tanb\lsim m_t/m_b$ in most MSSM models.

\subsection{The decoupling limit}
\medskip

In many models with extended Higgs sectors, a parameter regime exists
in which one Higgs boson is light (of order $m_Z$)
and all other Higgs scalars are very heavy $(\gg\mz$).
In this case, one can formally integrate out the heavy scalar states.  The
effective low-energy Higgs theory is precisely that of the SM Higgs
boson.  This is called the decoupling limit~\cite{decoupling}.  
The MSSM exhibits the
decoupling limit for $\mha\gg\mz$.  In this case, it is easy to verify
that $\mha\simeq\mhh
\simeq\mhpm$, up to corrections of ${\cal O}(\mz^2/\mha)$, and
$\cos(\beta-\alpha)=0$ up to corrections of ${\cal O}(\mz^2/\mha^2)$.
Examining the Higgs couplings listed above, one can check that
in the limit of $\cos(\beta-\alpha)\to 0$,
all $h^0$ couplings to SM particles approach their SM limits,
as expected.

\section{Present status of the Higgs boson}
\bigskip

\subsection{Direct Higgs mass bounds}
\medskip

From 1989--2000, experiments
at LEP searched for $e^+e^-\to Z\to h^0Z$ (where one of the $Z$-bosons
is on-shell and one is off-shell).  No significant evidence was found
leading to a lower bound on the Higgs mass,
$m_h> 114.4~{\rm GeV}$ at $95\%~{\rm CL}$~\cite{LEPHIGGS}.
Recent data from the Tevatron extends the disallowed Higgs mass
region by excluding $158~{\rm GeV}<m_h<175~{\rm GeV}$
at $95\%~{\rm CL}$~\cite{TEVHIGGS}.

The MSSM Higgs mass bounds are more complicated, since they depend
on a variety of MSSM parameters.  Since supersymmetric
radiative corrections to the lightest MSSM Higgs mass
must be significant to avoid exclusion, the MSSM Higgs mass bounds
depend sensitively on the multi-dimensional MSSM parameter space.
In the maximal-mixing scenario, the LEP Higgs search rules out
(at 95\% CL) charged Higgs bosons with $\mhpm< 79.3$~GeV 
and neutral Higgs masses with  $\mhl< 92.9$~GeV and 
$\mha<93.4$~GeV~\cite{mssmhiggslimits}.
In certain other scenarios, it is possible to significantly relax
these bounds.  For example, in the CPX scenario~\cite{CPX}, supersymmetric
radiative corrections introduce CP-violating phenomena into the MSSM
Higgs sector, in which case an arbitrarily light Higgs boson is still
allowed for $3\lsim \tanb\lsim 10$~\cite{CPXLEP}.

\subsection{Indirect Higgs mass bounds}
\medskip

Precision electroweak data puts indirect bounds on the Higgs mass.
One can fit a plethora of electroweak observables, based on the
predictions of the SM, in which the Higgs mass is allowed to float.
The global fit of precision electroweak data yields a
$\chi^2$ distribution for the goodness of fit as a function of the
Higgs mass.  If the direct Higgs search bounds are omitted, the
favored central value of the Higgs mass obtained by the LEP
Electroweak Working group is
89 GeV, with a one-sigma experimental uncertainty of $+35$ and 
$-26$~GeV~\cite{LEPEWWG}.
The corresponding one-sided 95\% CL upper bound 
excludes $m_h<158$~GeV.
Including the LEP Higgs search data, the upper bound increases to 185
GeV.  Similar results are obtained by the \texttt{GFITTER} collaboration, which
quotes~\cite{GFITTER}:
\beq
m_h=120.6^{+17.0[+34.3]}_{-5.2[-6.2]}~{\rm GeV}\,,
\eeq
based on \text{all} direct and indirect data, where both 1$\sigma$ and
2$\sigma$ error bars are given (the latter in the square brackets).
The minimum $\chi^2$ associated with the \texttt{GFITTER} fit is $\chi^2_{\rm
  min}=17.82$ for 14 degrees of freedom.

%When precision electroweak data is confronted with the MSSM Higgs
%sector, the following conclusions ensue.  (1) In the 
%decoupling limit (assuming that the SUSY particles are somewhat heavy),
%the effects of the heavy Higgs states and the SUSY particles decouple and
%the global SM fit applies.
%(2)
%In the latter case, $h^0$ is a SM-like Higgs boson whose mass lies below
%about $130$~GeV in the \textit{preferred} Higgs mass range!
%(3)
%If SUSY particle masses are not too heavy, they can have small
%effects on the fit to precision electroweak data.  With additional
%degrees of freedom, the goodness of fit can be slightly improved
%(and possibly argue for SUSY masses close to their present
%experimental limits).
%(4)
%The MSSM fit is further improved if one wishes to ascribe deviations of
%$(g-2)_\mu$ from their SM expectations to the effects of
%superpartners.

\subsection{Can a light Higgs boson be avoided?}
\medskip

\begin{figure}[b!]
\begin{center}
\includegraphics*[width=7.9cm]{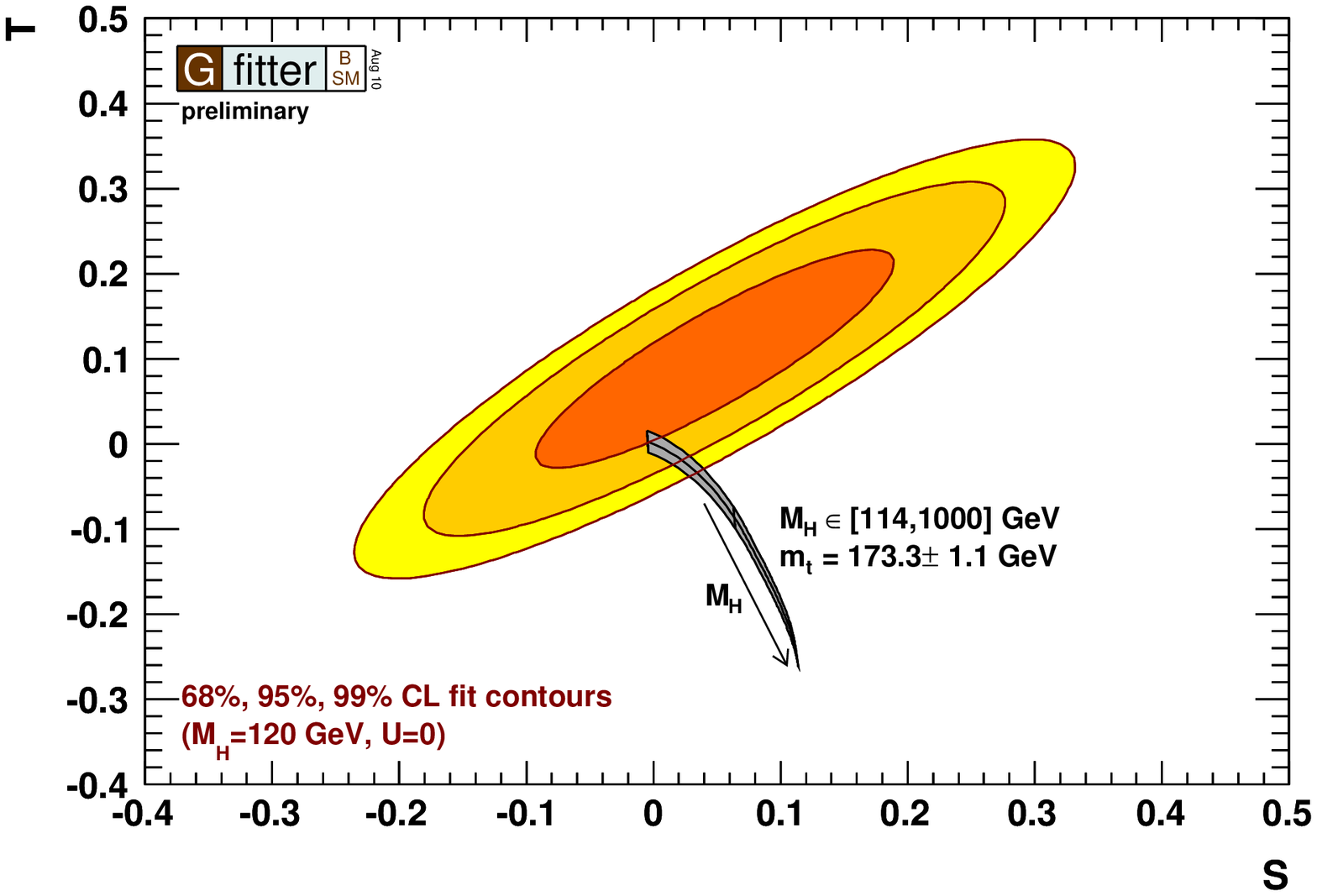}
\includegraphics*[width=7.9cm]{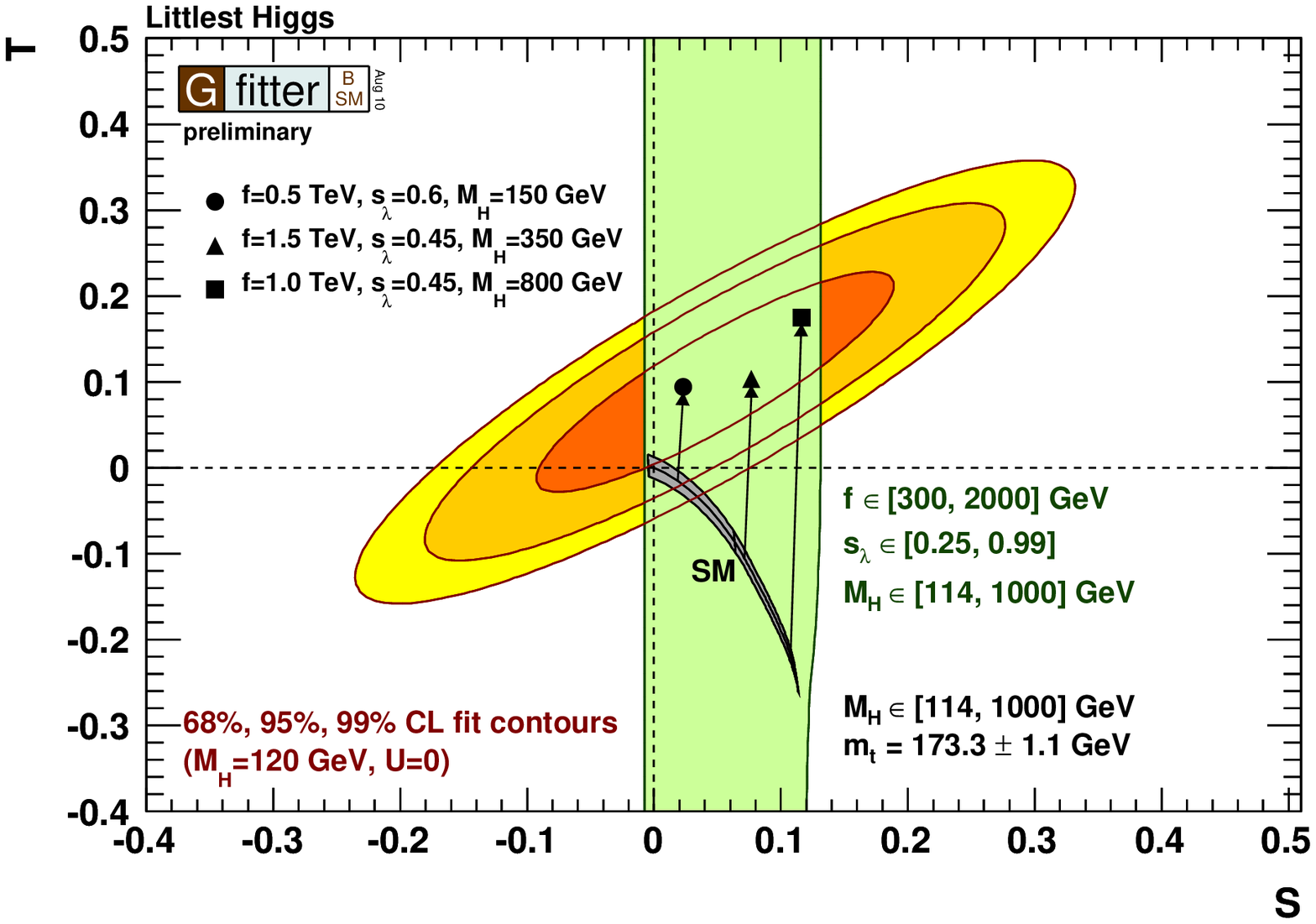}
\end{center}
\caption{(a) In the left panel, the 68\%, 95\% and 99\% CL contours
in the $S$ vs. $T$ plane based on the \texttt{GFITTER} global Standard 
Model fit to electroweak precision data.  The point $(S,T)=(0,0)$ 
corresponds to $m_t=173.1$~GeV and $m_h=120$~GeV.
(b) In the right panel, the \texttt{GFITTER} fit to electroweak precision data
is exhibited for the littlest Higgs model of \Ref{lhiggs} 
(with $T$-parity) for the
parameters shown.  Heavier Higgs masses are allowed, if the
effects of the Higgs bosons to the gauge boson vacuum polarization
functions are compensated by the effects due to new particles 
and interactions contained in the littlest Higgs
model.  Taken from \Ref{GFITTER}. \label{st}}
\end{figure}

If new physics beyond the Standard Model (SM) exists, 
%it almost certainly couples to $W$ and $Z$ bosons.  Then, 
then one expects shifts in
the $W$ and~$Z$ masses due to new contributions to the one-loop corrections.
In many cases, these effects can be parameterized in terms of two
quantities, $S$ and~$T$ introduced by Peskin and Takeuchi~\cite{peskin}:
\beqa
\overline\alpha\, T &\equiv& \frac{\Pi^{\rm new}_{WW}(0)}{m_W^2}
-\frac{\Pi^{\rm new}_{ZZ}(0)}{m_Z^2}\,,  \\
\frac{\overline\alpha}{4\overline s^2_Z\overline c^2_Z}\,S &\equiv &
\frac{\Pi^{\rm new}_{ZZ}(m_Z^2)-\Pi^{\rm new}_{ZZ}(0)}{m_Z^2}
-\left(\frac{\overline c^2_Z-\overline s^2_Z}{\overline c_Z\overline s_Z}
\right)\frac{\Pi^{\rm new}_{Z\gamma}(m_Z^2)}{m_Z^2}
-\frac{\Pi^{\rm new}_{\gamma\gamma}(m_Z^2)}{m_Z^2}\,,
\eeqa
where $s\equiv \sin\theta_W$, $c\equiv \cos\theta_W$, and barred
quantities are defined in the $\overline{\rm MS}$ scheme evaluated
at $m_Z$.  The $\Pi_{V_a V_b}^{\rm new}(q^2)$ are the new physics contributions
to the one-loop gauge boson vacuum polarization functions.
The region in the $S$--$T$ plane consistent with precision electroweak data
is shown in Fig.~\ref{st}(a).

In order to avoid the conclusion of a
light Higgs boson, new physics beyond the SM must enter
at an energy scale between 100 GeV and 1 TeV.
This new physics can be detected at~the~LHC
either through direct observation of new physics beyond the SM
or by improved precision measurements at future colliders
that can detect small deviations
from SM predictions.  Although the precision electroweak data is 
suggestive of a light Higgs boson,
one cannot definitively rule out a significantly heavier Higgs boson
if the new physics conspires to yield a total contribution to $S$ and $T$
that is consistent with the precision electroweak data
[e.g., see Fig.~\ref{st}(b)].
%although the measured $S$ and $T$ impose strong constraints on
%alternative approaches.

\section{Prospects for Higgs discovery at the Tevatron}
\bigskip

\begin{figure}[ht!]
\begin{center}
\includegraphics*[width=8cm,angle=-90]{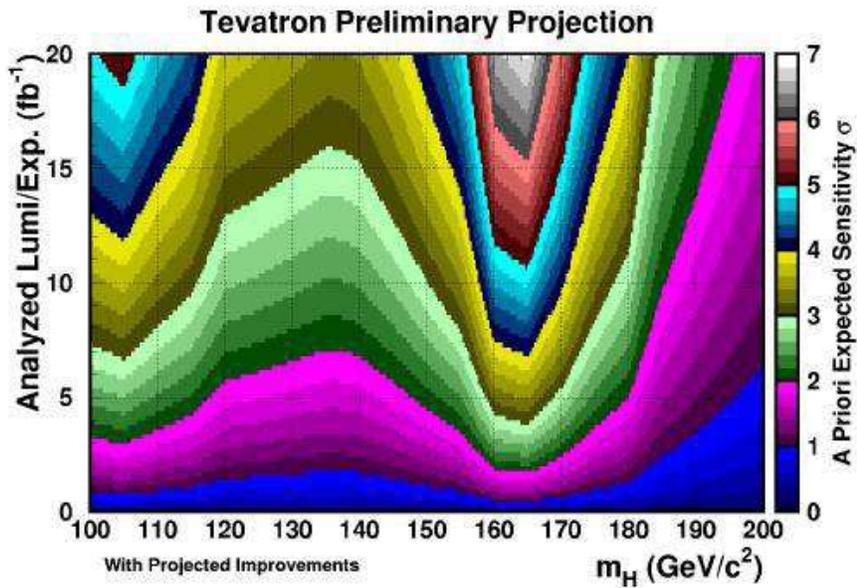}
\end{center}
\caption{\label{tevatron} Higgs Boson sensitivity with projected
  improvements per experiment~\cite{Tevproject}.}
\end{figure}

The Tevatron will continue to take data through the end of 2011.
In addition to an increased integrated luminosity, there is still room 
for some improvements in the Higgs search analysis.  The possibility
of extending the Tevatron run by another three years is currently 
under discussion.  To facilitate this discussion, a collaboration of
Fermilab theorists and experimentalists produced a white paper making the
case for an extended Run III of the Tevatron from the end of 2011
through 2014~\cite{Tevproject}.  
The projected improvement of the Tevatron Higgs search
is shown in Fig.~\ref{tevatron}.

\noindent
In Table 1, the expected sensitivity
of the Tevatron Higgs search for Higgs masses of 115 GeV, 130 GeV
and 145 GeV as a function of the analyzable luminosity per
experiment is given.
\begin{center}
\begin{table}[ht]
\centering
\begin{tabular}{|c|ccc|}
\hline
Analyzable Lum/Exp & 115~GeV & 130~GeV & 145~GeV\\
\hline
5~fb$^{-1}$ & $2.2~\sigma$ & $1.7~\sigma$ & $1.9~\sigma$\\
10~fb$^{-1}$ & $3.1~\sigma$ & $2.5~\sigma$ & $2.7~\sigma$\\
15~fb$^{-1}$ & $3.8~\sigma$ & $3.0~\sigma$ & $3.2~\sigma$\\
20~fb$^{-1}$ & $4.4~\sigma$ & $3.5~\sigma$ & $3.7~\sigma$\\
\hline
\end{tabular}
\caption{Sensitivity to the Standard Model Higgs Boson combining all
modes.  The low mass $\leq 130$~GeV mode is principally $q\bar{q}\to
(W,Z)+(h\to b\bar{b})$; the higher mass $\geq 130$~GeV mode is
principally $gg\to h\to WW^*$.  Taken from \Ref{Tevproject}.} 
\end{table}
\end{center}

\section{Higgs phenomenology at the LHC}
\bigskip

Although it is possible that evidence for the Higgs boson may
emerge from future Tevatron running, the discovery of the Higgs
boson and the identification of its properties are expected to 
take place at the LHC.  Once the Higgs boson is discovered,
a program of Higgs physics at the LHC must address the following
important questions:
\begin{itemize}
\item
How many Higgs states are there?
\item
Assuming one Higgs-like state is discovered,
\begin{itemize}
\item
is it a Higgs boson?
\item
is it \textit{the} SM Higgs boson?
\end{itemize}
\end{itemize}
The measurement of Higgs boson properties will be critical
in order to answer the last two questions:
\begin{itemize}
\item
mass, width, CP-quantum numbers (is the Higgs sector CP-violating?);
\item
branching ratios and Higgs couplings;
\item
reconstructing the Higgs potential (Higgs self-couplings).
\end{itemize}

\subsection{Mechanisms for SM Higgs production at the LHC}
\medskip

\begin{figure}[ht!]
\begin{center}
\includegraphics[width=3cm]{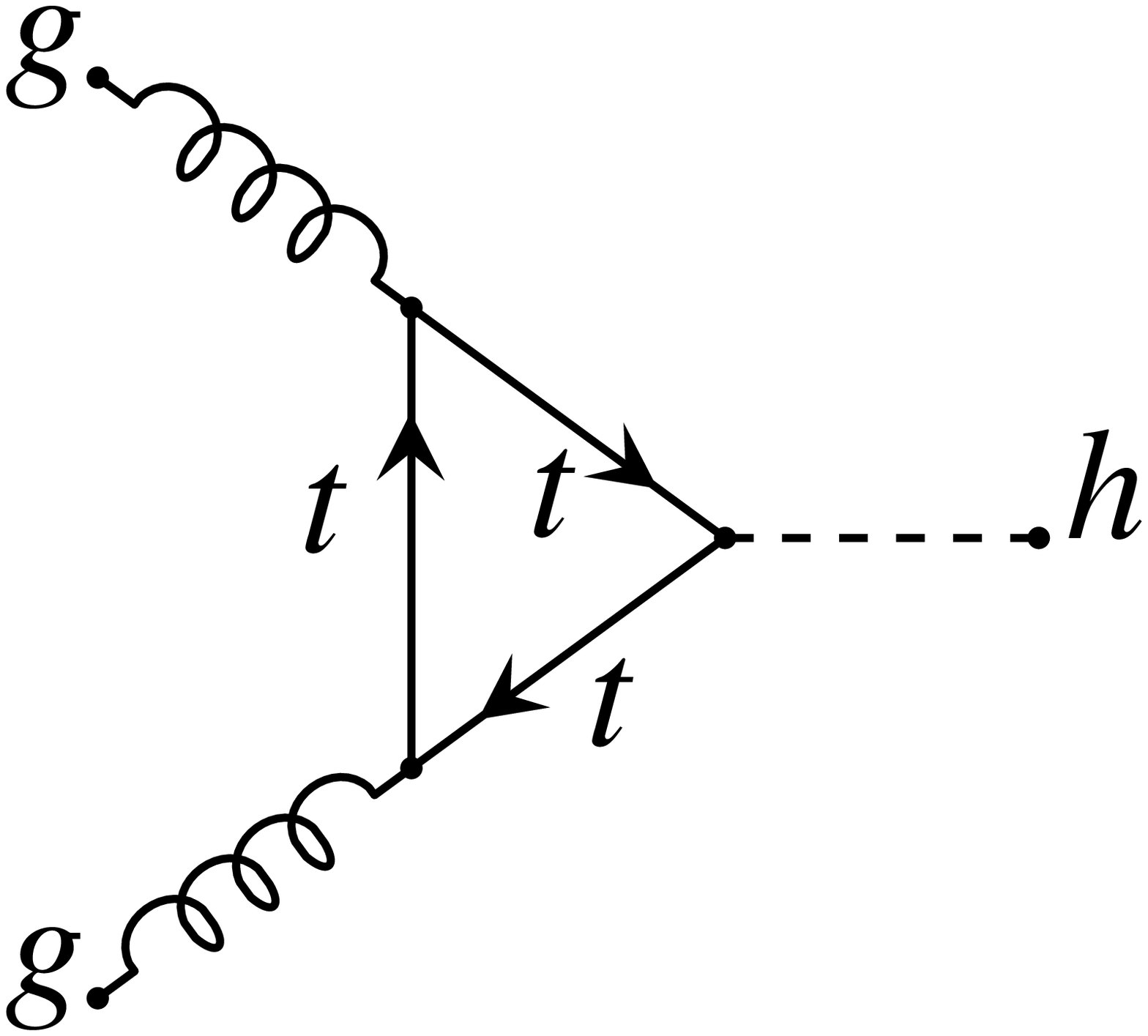}\hspace{0.7cm}
\includegraphics[width=3cm]{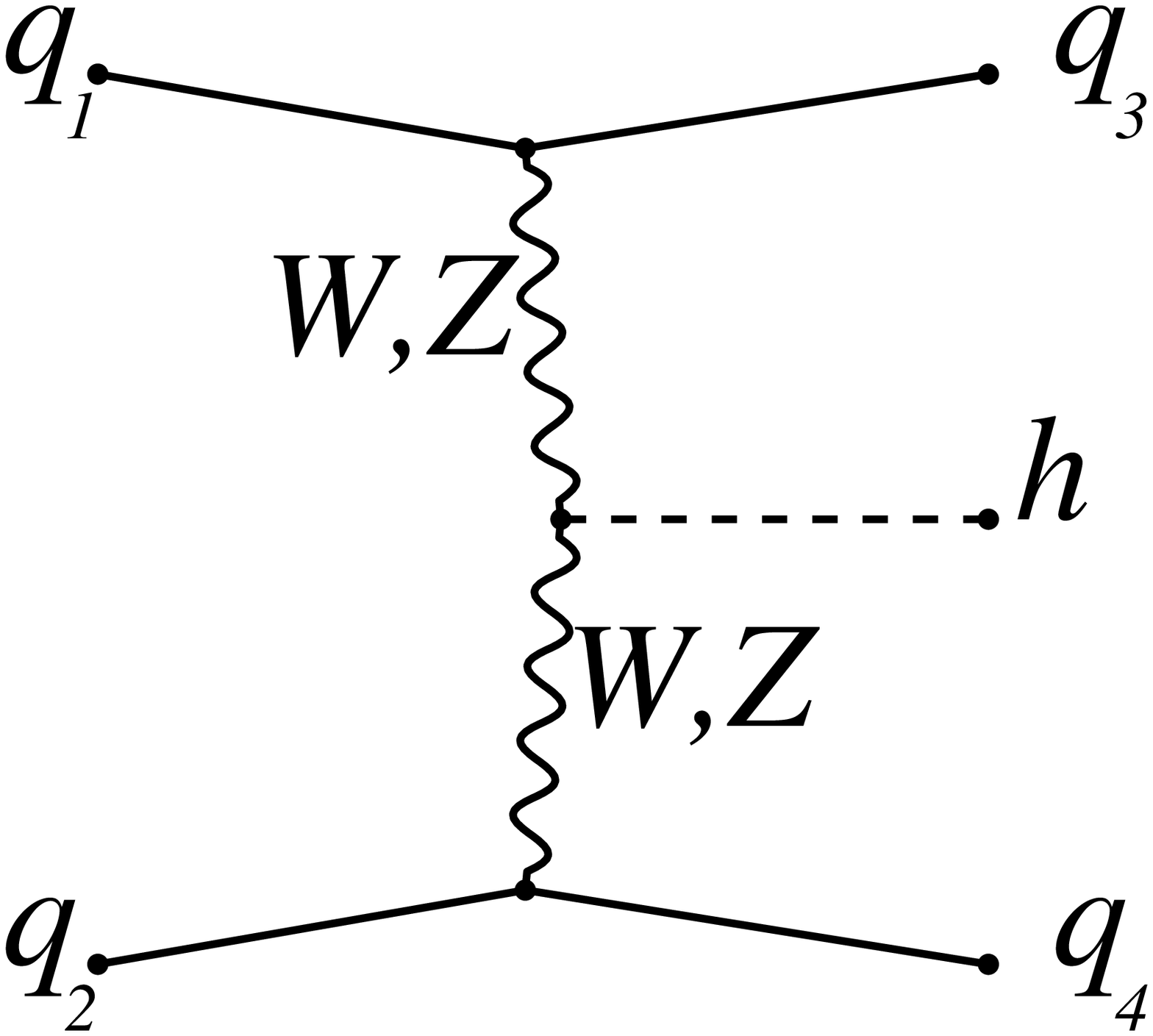}\hspace{0.7cm}
\includegraphics[width=3.5cm]{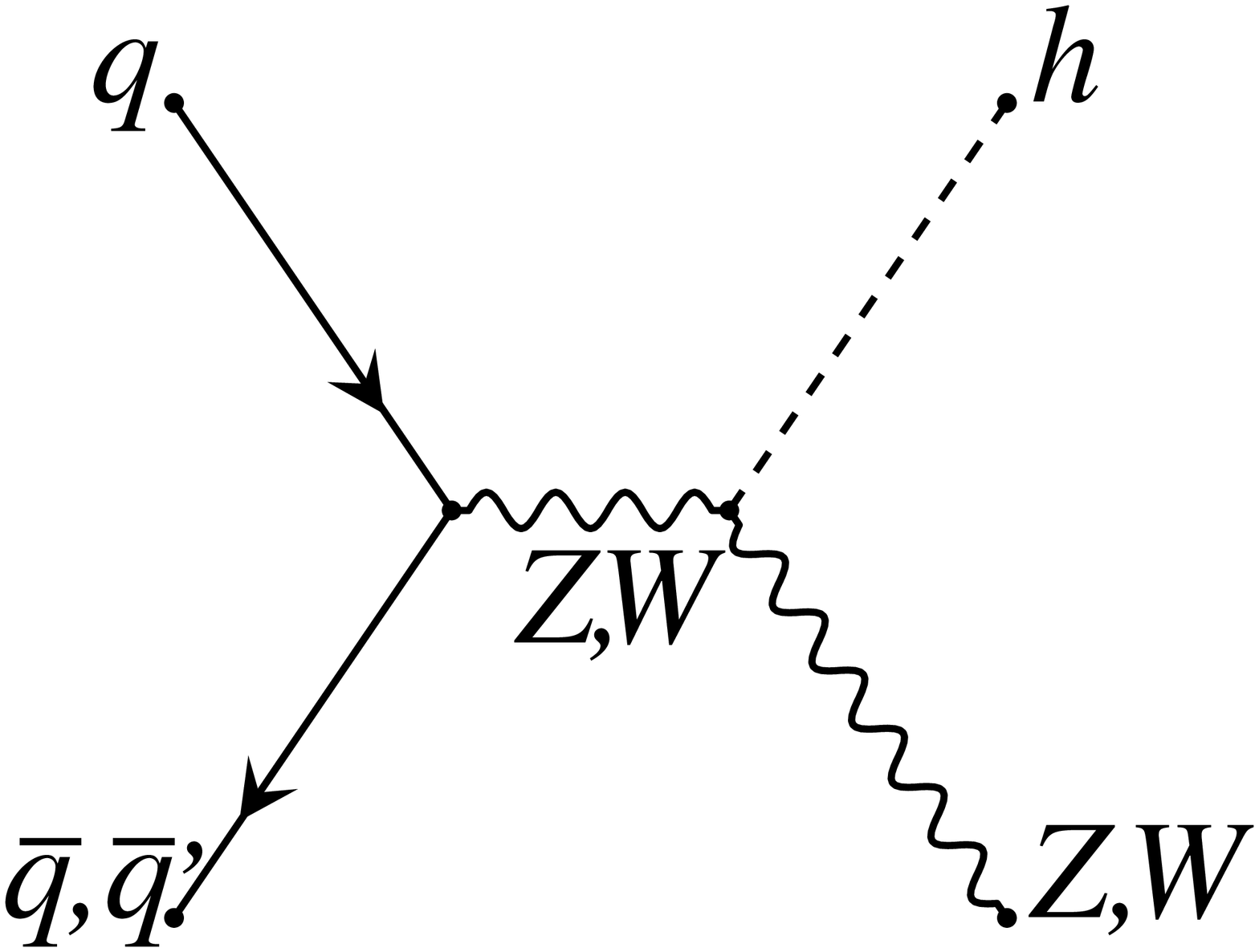}\hspace{0.7cm}
\includegraphics[width=3cm]{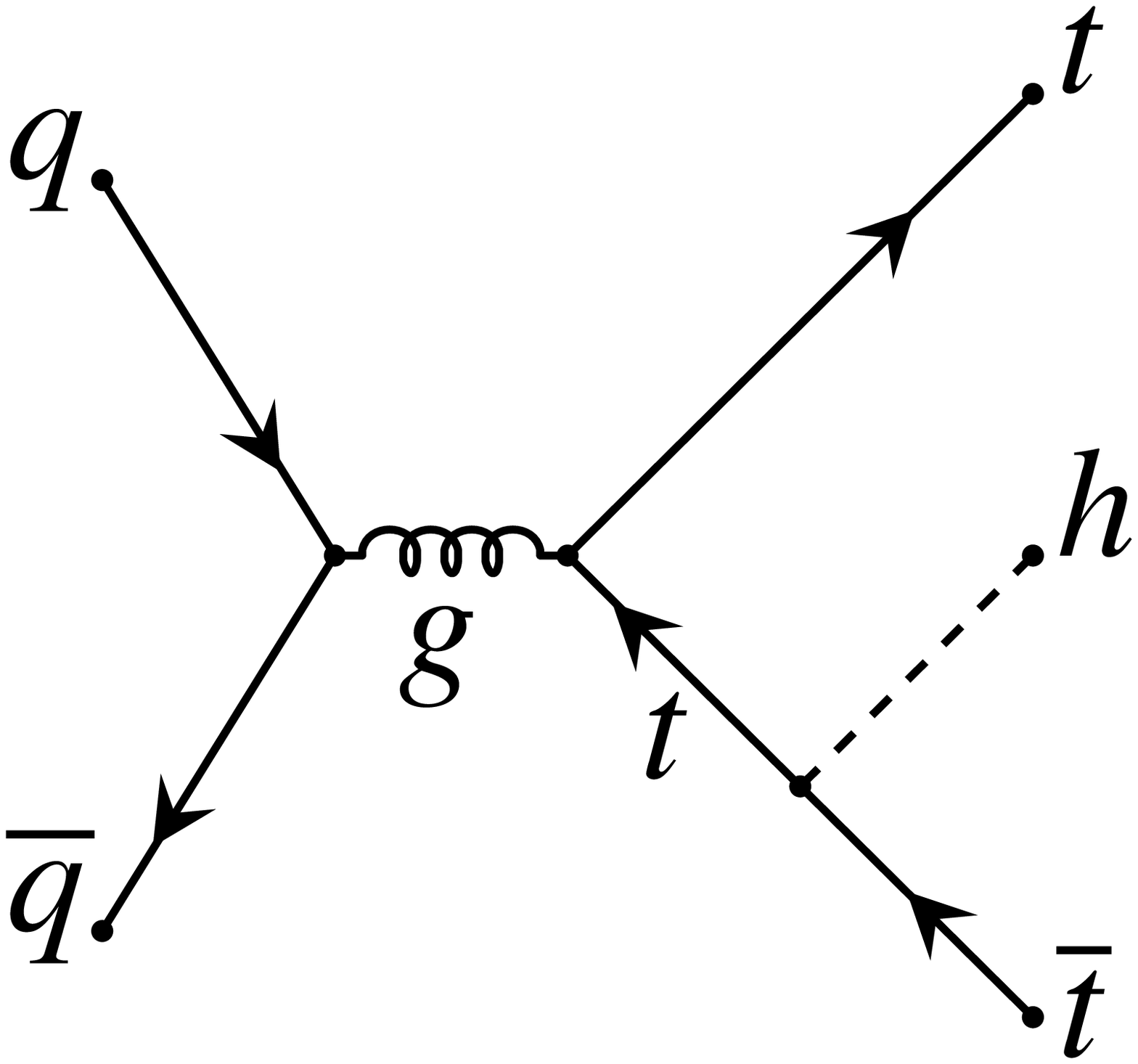}
\end{center}
\caption{\label{hprod} SM Higgs production mechanisms at 
hadron colliders }
\end{figure}

At hadron colliders, the relevant Higgs production processes
are shown in Fig.~\ref{hprod}.

The Higgs boson is then detected via its decay products.
The dominant channels for observing the Higgs boson at the LHC are:
\beqa
&& gg\to\hsm\,,\quad \hsm\to\gamma\gamma\,,\,
VV^{(*)}\,, \nonumber  \\
&& qq\to qqV^{(*)} V^{(*)}\to qq\hsm,\quad \hsm\to\gamma\gamma,\,
\tau^+\tau^-, \,VV^{(*)}\,, \nonumber \\
&& q\bar q^{(\prime)}\to V^{(*)}\to V\hsm\,, \quad \hsm\to b\bar b\,,
WW^{(*)}\,,\nonumber \\
&& gg, q\bar q\to t\bar t\hsm, \quad \hsm\to b\bar b,
\,\gamma\gamma, \,WW^{(*)}\,.\nonumber
\eeqa
where $V=W~{\rm or}~Z$ is a physical gauge boson and $V^*$ is 
a virtual (off-mass-shell) gauge boson.

\subsection{SM Higgs cross sections and branching ratios}
\medskip

\begin{figure}[ht!]
\begin{center}
\includegraphics*[width=7.4cm,angle=-90]{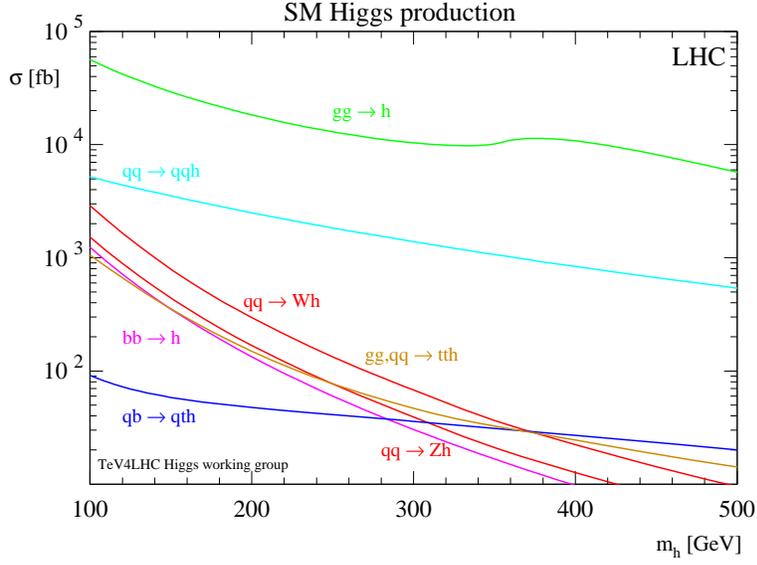}
\end{center}
\caption{\label{hxsec}  SM Higgs production cross-sections 
for $\sqrt{s}=14$~TeV at the LHC (taken from \Ref{xsec}).}
\end{figure}

\begin{figure}[ht!]
\begin{center}
%\vspace{0.1in}
\resizebox{\textwidth}{!}{
\includegraphics*{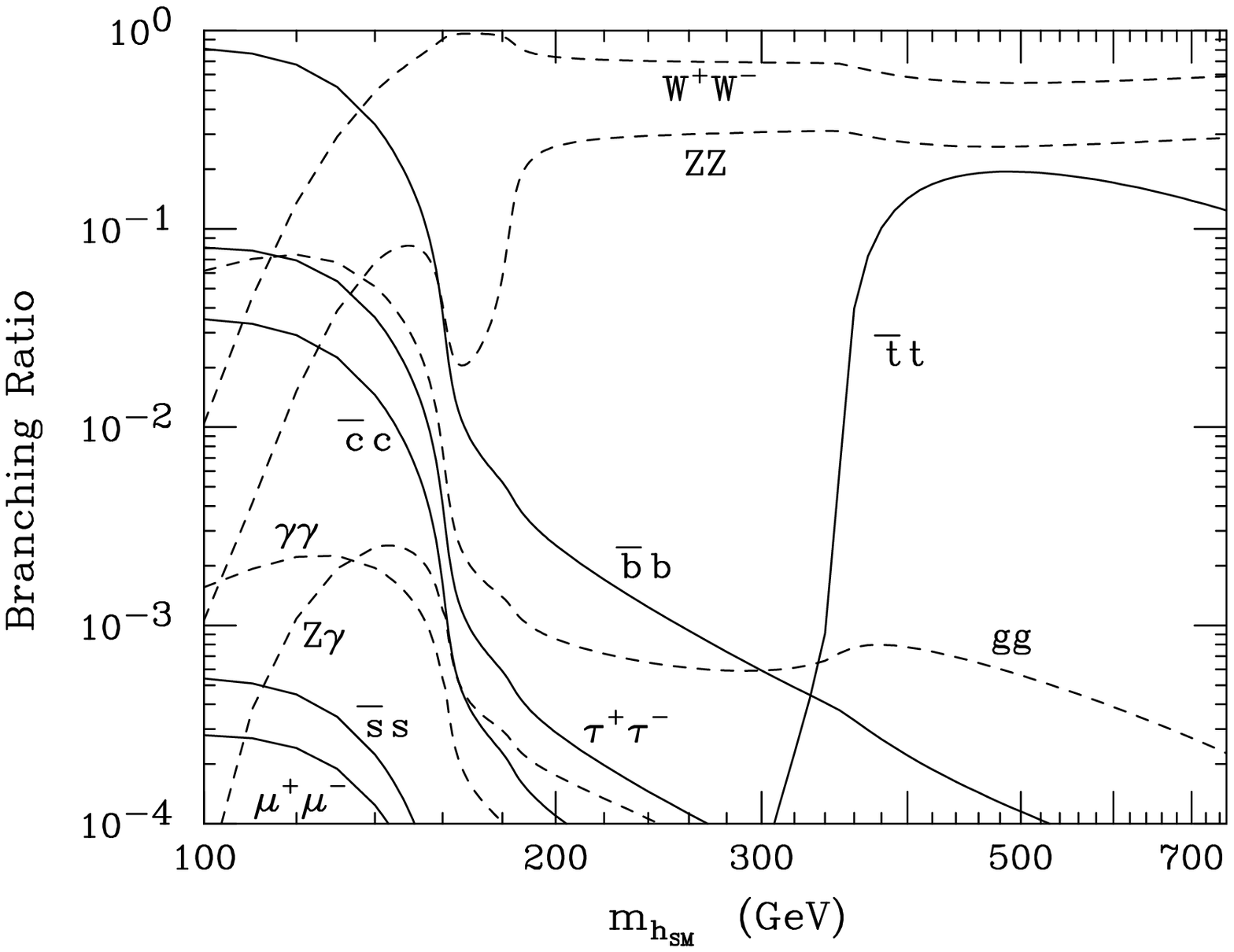}
\hspace*{3mm}
\includegraphics*{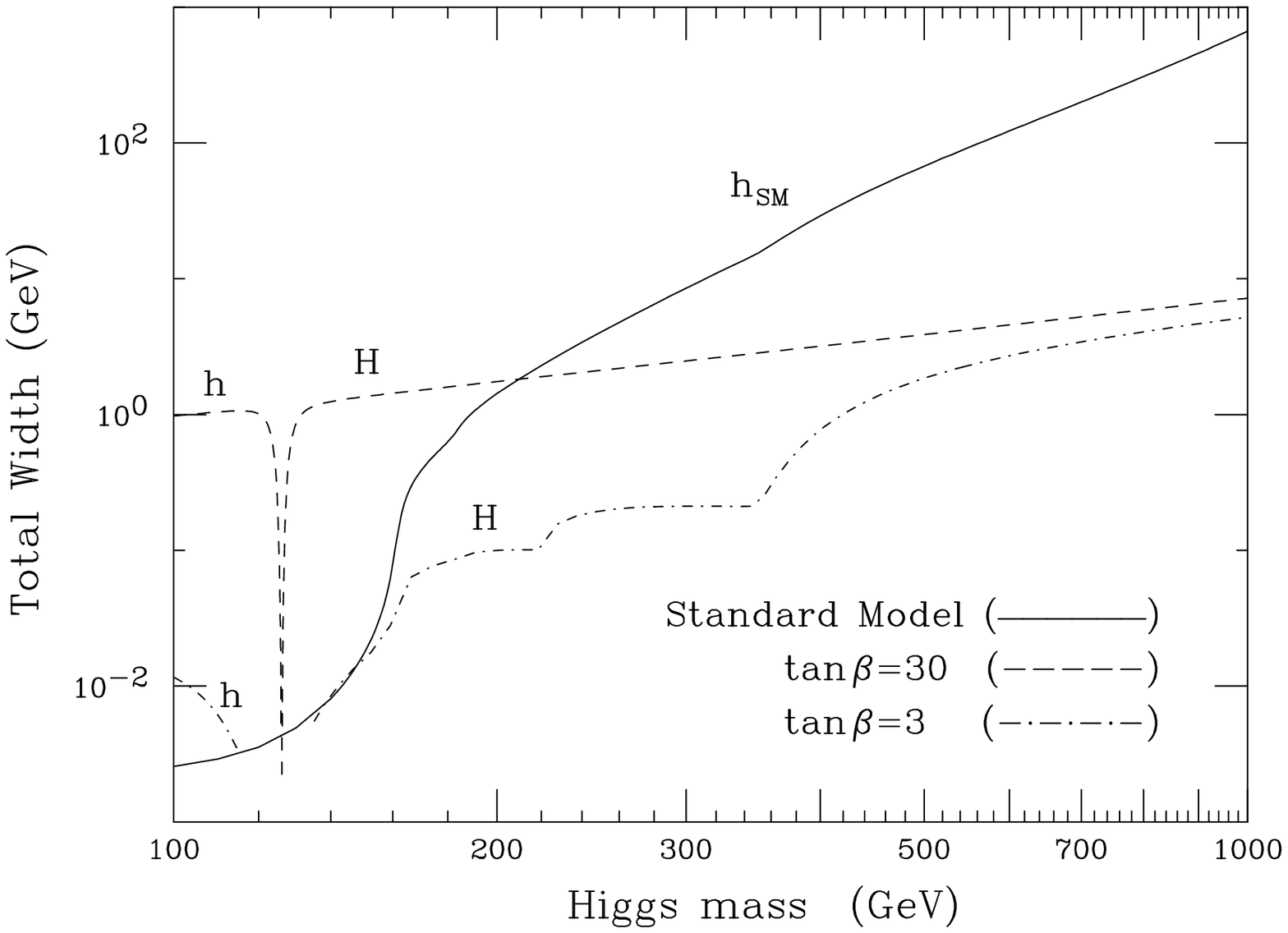}
}
\end{center}
\caption{\label{hbr} (a) In the left panel, the branching ratios
of the SM Higgs boson are shown as a function of the Higgs mass.
Two-boson [fermion-antifermion]
final states are exhibited by solid [dashed] lines.
(b) In the right panel, the total width of the Standard Model
Higgs boson (denoted by $h_{\rm SM}$) 
is shown as a function of its mass.  For comparison, 
the widths of the two CP-even scalars, $\hl$ and $\hh$ of the
MSSM are exhibited 
for two different choices of MSSM parameters ($\tan\beta=3$ and
30 in the maximal mixing scenario; the onset of the $\hh\to\hl\hl$ and
$\hh\to t\bar t$ thresholds in the $\tan\beta=3$ curve are clearly evident).
Taken from \Ref{carenahaber}.}
\end{figure}

The SM Higgs production cross-sections for $\sqrt{s}=14$~TeV at the
LHC are shown in Fig.~\ref{hxsec}.
The discovery channels for Higgs production depend
critically on the Higgs branching ratios.  The branching ratios 
and total width of the SM Higgs boson are shown in
Fig.~\ref{hbr}. 

\subsection{LHC prospects for SM Higgs discovery}
\medskip

An examination of Fig.~\ref{hbr} indicates
that for $m_h<135$~GeV, the decay
$h\to b\bar{b}$ is dominant, whereas for $m_h>135$~GeV, the decay
$h\to WW^{(*)}$ is dominant (where one of the $W$ bosons must be
virtual if $m_h<2\mw$).  These two Higgs mass regimes require
different search strategies.  For the lower mass Higgs scenario,
gluon-gluon fusion to the Higgs boson followed by $h\to 
b\bar{b}$ cannot be detected as this signal is overwhelmed by
QCD two-jet backgrounds.  Instead, the
Tevatron Higgs search employs associated
production of the Higgs boson with a $W$ or $Z$ 
followed~by~$h\to b\bar{b}$.  
This process is more difficult at the LHC,
where the signal-to-background discrimination is more severe.
Instead, the LHC relies on the rare decay mode, $h\to\gamma\gamma$,
which has a branching ratio of about $2\times 10^{-3}$ in the Higgs
mass regime of interest.  In the higher Higgs mass regime, both the Tevatron
and the LHC rely primarily on $h\to WW^{(*)}$ for Higgs masses below 
200~GeV.  Should the Higgs mass be significantly larger than 
the current expectations based on precision electroweak data,
then the LHC can discover the Higgs boson through the so-called golden
channel, $h\to ZZ\to \ell^+\ell^-\ell^+\ell^-$.
\begin{figure}[ht!]
\begin{center}
\includegraphics*[width=19.5cm]{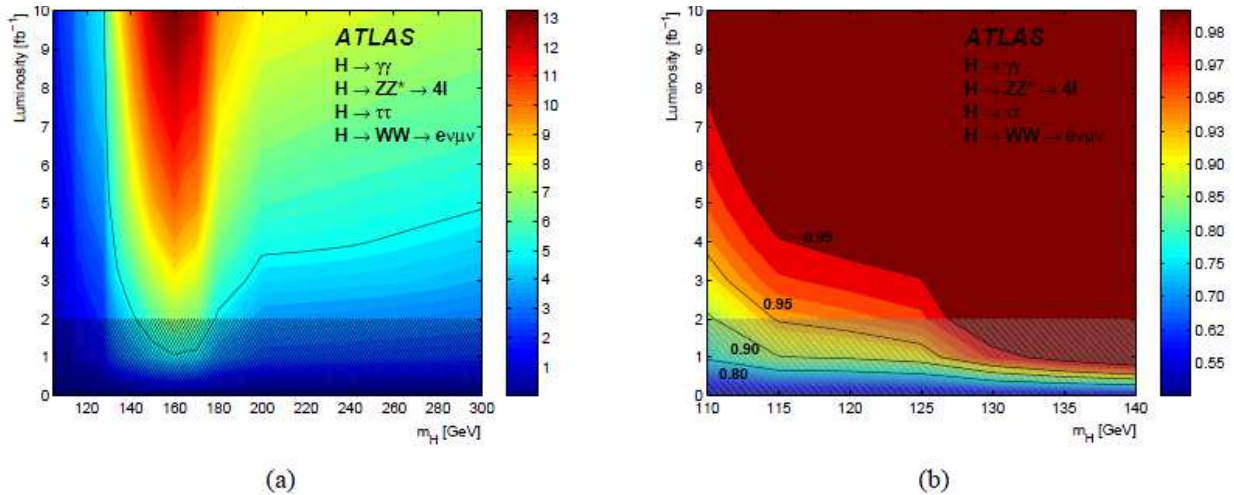}
\end{center}
\caption{\label{atlashiggs} (a) The left panel depicts the 
significance contours for different SM
  Higgs masses and integrated luminosities.  The thick curve
  represents the $5\sigma$ discovery contour.  The median
  significance is shown with a color according to the legend. 
 (b) The right panel depicts the expected luminosity
  required to exclude a Higgs boson with a mass $m_H$ at a confidence
  level given by the corresponding color.   In  both panels,  the
  hatched area below 2~fb$^{-1}$ indicates the region where the
  approximations used in the combination are not accurate, although
  they are expected to be conservative.  Taken from \Ref{atlash}.} 
\end{figure}

\begin{figure}[ht!]
\begin{center}
\includegraphics*[width=7.66cm]{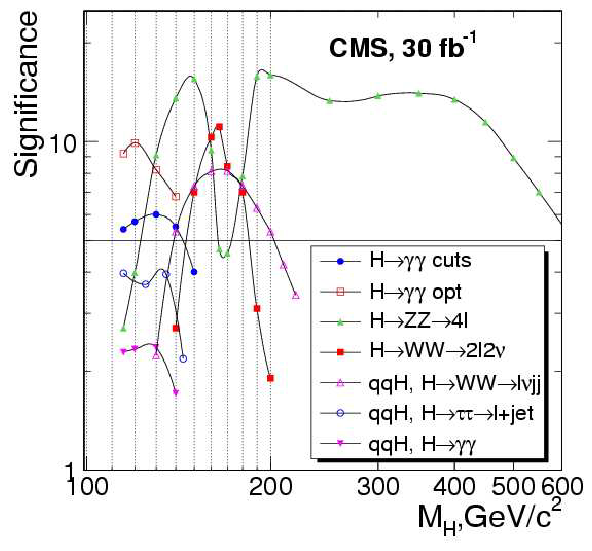}
\hspace{1mm}
\includegraphics*[width=7.96cm]{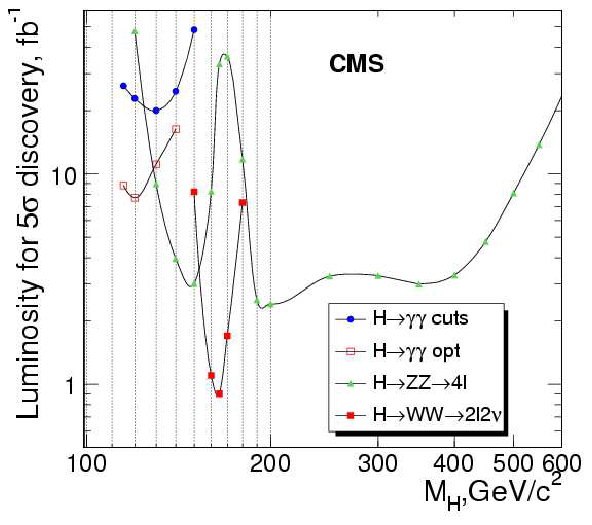}
\end{center}
\caption{\label{cmshiggs} (a) The left panel depicts the 
signal significance as a function of the SM Higgs boson mass for
30~fb$^{-1}$ at $\sqrt{s}=14$~TeV, for the different Higgs
boson production and decay channels.
 (b) The right panel depicts the integrated luminosity
  required for a $5\sigma$ discovery of the inclusive Higgs
boson production, $pp\to h+X$, with the Higgs boson decay modes
$h\to\gamma\gamma$,
$h\to ZZ\to \ell^+\ell^-\ell^+\ell^-$ and $h\to W^+W^-\to
\ell^+\nu_\ell\, \ell^-\bar{\nu}_\ell$.  Taken from \Ref{cmsh}.}
\end{figure}

The ATLAS projections~\cite{atlash} exhibited in Fig.~\ref{atlashiggs}
suggest
that in the Higgs mass range of $114~{\rm GeV}\lsim m_h\lsim 130$~GeV, the
discovery of the Higgs (primarily through the rare $h\to\gamma\gamma$
decay mode) will require an integrated luminosity larger than
$10~{\rm fb}^{-1}$ at $\sqrt{s}=14$~TeV.  The CMS projections~\cite{cmsh} 
shown in Fig.~\ref{cmshiggs} are slightly more optimistic
in the low mass Higgs regime based on an
``optimized'' neural net analysis of the $h\to\gamma\gamma$ mode.

\begin{figure}[ht!]
\begin{center}
\vspace{0.2in}
\includegraphics*[width=12.49cm]{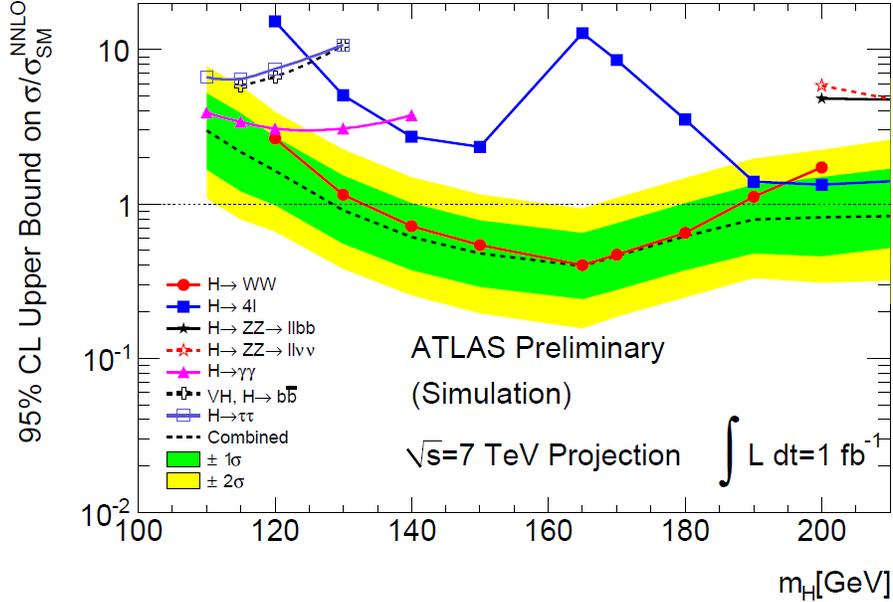}
\end{center}
\caption{\label{atlas7} 
The multiple of the cross-section of a SM Higgs boson that can be
excluded using $1~{\rm fb}^{-1}$ of data at $\sqrt{s}=7$~TeV.  
At each mass, every channel giving reporting on it is used.  The
green and yellow bands indicate the range in which the limit is
expected to lie, depending on the data.  Taken from \Ref{at7}}
\end{figure}

Currently, the LHC is running at half the design energy at
significantly lower luminosities than those displayed in
Figs.~\ref{atlashiggs} and \ref{cmshiggs}.  Current projections
suggest that at the end of 2011, the LHC running at $\sqrt{s}=7$~TeV
will accumulate a total integrated luminosity per
experiment of about 1~fb$^{-1}$.  Based on these projections, the ATLAS and
CMS Higgs searches will yield a $5\sigma$ discovery in only a limited
Higgs mass range, with a possible 95\% exlusion over a 
somewhat larger Higgs mass range~\cite{at7,cms7}.  As an example, 
Fig.~\ref{atlas7} depicts the $95\%$ Higgs mass
exclusion region anticipated at the end of the 2011 run~\cite{at7}.  These
results would likely match or even exceed the corresponding high mass
($m_h\geq 130$~GeV) exclusion range expected from the Tevatron.

\begin{table}[ht]
\begin{minipage}[b]{0.5\linewidth}\centering
\begin{tabular}{|c||
 c| c| c| c|}
\hline
$\mathbb{H}_{0} \Downarrow \mathbb{H}_{1}\Rightarrow$ & $0^{+}$ & $0^{-}$ & $1^{-}$ & $1^{+}$ \\
\hline
\hline
 $0^{+}$  & -- & 17 & 12 & 16 \\
%\hline
 $0^{-}$  & 14 & -- & 11 & 17 \\
%\hline
 $1^{-}$  & 11 & 11 & -- & 35 \\
%\hline
 $1^{+}$  & 17 & 18 & 34 & -- \\
\hline
\end{tabular}
\end{minipage}
\hspace{0.5cm}
\begin{minipage}[b]{0.5\linewidth}
\centering
\begin{tabular}{|c||
 c| c| c| c|}
\hline
$\mathbb{H}_{0} \Downarrow \mathbb{H}_{1}\Rightarrow$ & $0^{+}$ & $0^{-}$ & $1^{-}$ & $1^{+}$ \\
\hline
\hline
 $0^{+}$  & -- & 52  & 37 & 50 \\
%\hline
 $0^{-}$  & 44 & -- & 34 & 54  \\
%\hline
 $1^{-}$  & 33 & 32  & -- & 112  \\
%\hline
 $1^{+}$  & 54  & 55  & 109 & -- \\
\hline
\end{tabular}
\end{minipage}
\caption{Minimum number of observed events such that the median
significance for rejecting 
the null background-only hypothesis $\mathbb{H}_{0}$
in favor of the standard Higgs signal plus background
hypothesis $\mathbb{H}_{1}$ (assuming $\mathbb{H}_{1}$ is
right) exceeds $3\sigma$ and $5\sigma$, respectively, with $m_H$$=$$145$ GeV.
Based on an analysis of the $H\to ZZ^*$ decay mode.  Taken from
\Ref{lykken}.}
\end{table}

After the Higgs boson discovery, one must check that its properties
are consistent with the theoretical expectations.  Measurements of the
spin, parity and couplings to gauge bosons and fermions typically need
a large data sample (which requires tens to hundreds of fb$^{-1}$).
However, in some special circumstances, one can discriminate between
different hypotheses for spin and parity with a small data sample.
For example, for Higgs boson masses above about 130~GeV, 
it will be possible to isolate a data
sample of $h\to ZZ^{(*)}\to \ell^+\ell^-\ell^+\ell^-$.  By analyzing the
distributions and correlations of the five relevant angular variables
that describe the $\ell^+\ell^-\ell^+\ell^-$ events, one can achieve 
a median expected discrimination significance of $3\sigma$ with as few 
as 19 events for $m_h=200$~GeV and even better discrimination for the 
off-shell decays of an $m_h=145$~GeV Higgs boson~\cite{lykken}, as
shown in Table~2.  Thus, if a Higgs boson is discovered at the LHC
with $1~{\rm fb}^{-1}$, then it may already be possible to confirm a
preference for the expected $0^+$ spin-parity assignment with a small sample of
Higgs events. 

\subsection{LHC prospects for MSSM Higgs discovery}
\medskip

In the decoupling limit, the 
properties of the lightest CP-even Higgs boson of the MSSM
(denoted by $h^0$) are nearly identical to that of the SM Higgs boson.
Thus, all the SM Higgs search techniques apply without modification in
the MSSM.  Moreover, the other MSSM Higgs bosons, $H^\pm$, $\hh$ and $\ha$,
which are significantly heavier than $m_h$ (and $m_Z$), are roughly
degenerate in mass (cf. Section 2.4).  In the opposite limit (far from
the decoupling limit), all physical MSSM Higgs bosons are
typically below 200~GeV in mass and no single Higgs state possesses
couplings that precisely match those of the SM Higgs boson~\cite{intense}.

\begin{figure}[t!]
\begin{center}
\resizebox{\textwidth}{!}{
\scalebox{1.0}[1.25]{
\includegraphics*[width=10.8cm]{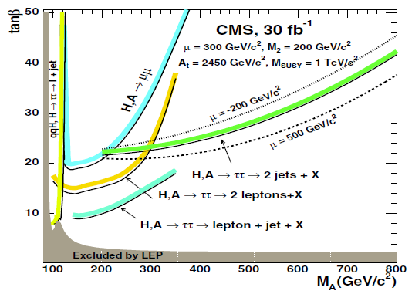}}
\hspace*{1mm}
\includegraphics*[width=12.5cm]{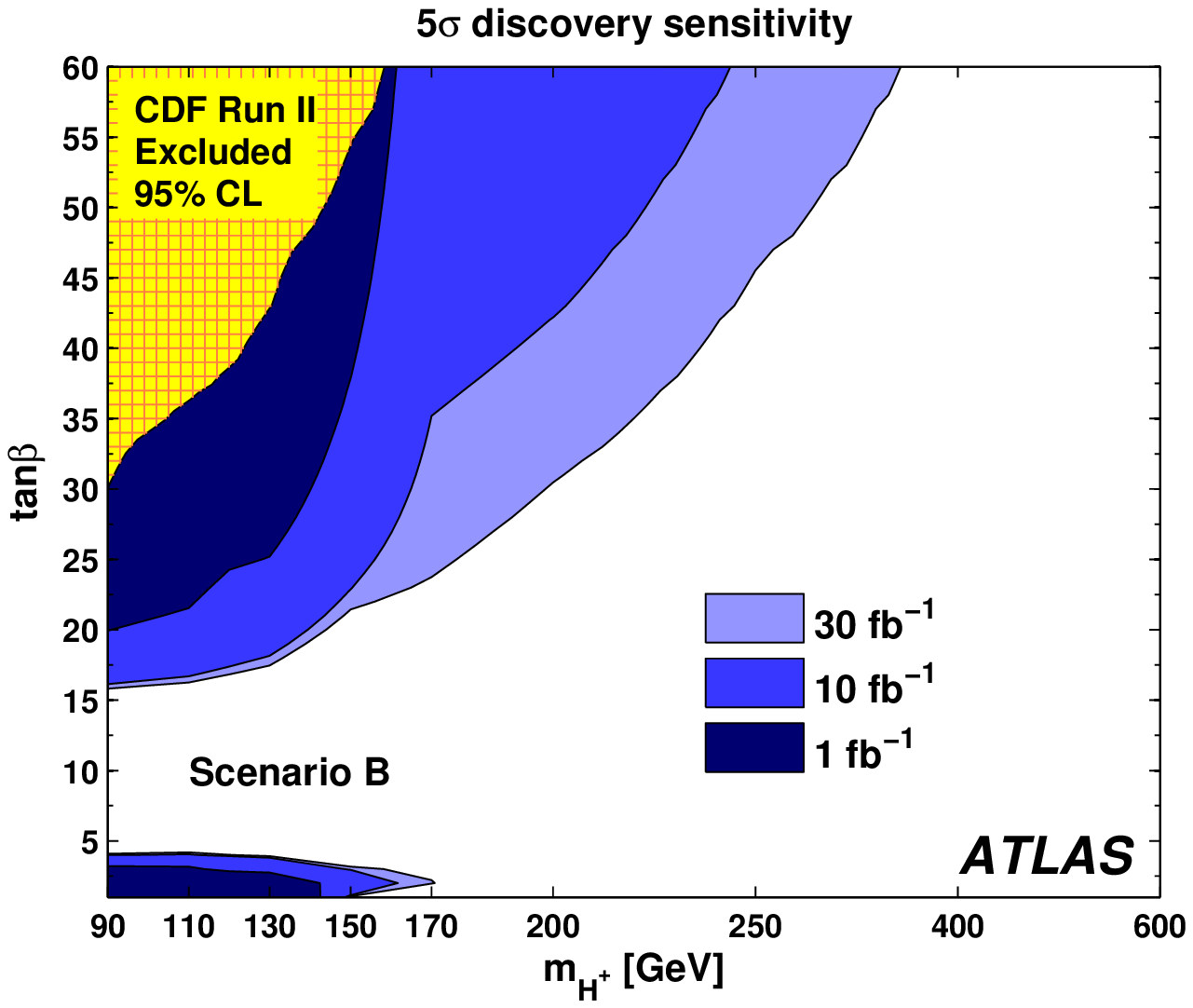}
}
\end{center}
\caption{In the left panel, the $5\sigma$ discovery potential 
in the maximal mixing scenario for the heavy
$\hh$ and $\ha$ scalars at CMS is shown for $\sqrt{s}=14$~TeV
and 30~fb$^{-1}$ of data (taken from \Ref{Abdullin:2005yn}).  
In the right panel, the $5\sigma$
discovery sensitivity for the charged Higgs mass is shown as a
function of $m_{H^\pm}$ and $\tan\beta$ (taken 
from \Ref{atlash}).\label{wedge}} 
\end{figure}

Cross-sections and branching ratios for the MSSM Higgs boson (which
depend on $\tanb$ as well as the corresponding Higgs mass) can be
found in refs.~\cite{carenahaber,djouadi2}.  Here, we simply note that:
\begin{itemize}
\item
gluon-gluon fusion can produce both neutral CP-even and CP-odd Higgs bosons.
\item
$VV$ fusion ($V=W$ or $Z$) can produce only CP-even Higgs bosons
(at tree-level).  Moreover, in the decoupling limit, the heavy CP-even
Higgs boson is nearly decoupled from the $VV$ channel.
\item
Neutral Higgs bosons can be produced in association with
$b\bar b$ and with $t\bar t$ in gluon-gluon scattering.
\item
Charged Higgs bosons can be produced in association with $t\bar b$ in
gluon-gluon scattering.
\item
If $m_{H^\pm}<m_t-m_b$, then $t\to bH^-$ is
an allowed decay, and the dominant $H^\pm$ production mechanism is
via $t\bar t$ production.
\item
Higgs bosons can be produced in pairs (e.g., $H^+ H^-$, $H^\pm\hl$, $\hl\ha$).
\item
Higgs bosons can be produced in cascade decays of SUSY particles.
\item
The total width of any one of the MSSM Higgs bosons is always $\lsim 1\%$ of 
its mass [cf.~Fig.~5(b)].
\item
Higgs search strategies depend on the location of the model in the 
$\mha$--$\tan\beta$ plane.
\end{itemize}
Simulations by the ATLAS and CMS collaborations shown in
Fig.~\ref{wedge} exhibit which
regions of the $\mha$--$\tanb$ plane are sensitive to the MSSM Higgs
search.  In particular, for large values of $\mha\gsim 200$~GeV
(corresponding to the decoupling limit), there is no sensitivity for
the discovery of the heavy MSSM Higgs states, $H^\pm$, $\hh$ and $\ha$,
for moderate values of $\tanb$.  This is the infamous ``LHC wedge''
region, in which the $\hl$ of the MSSM (whose properties are
nearly indistinguishable from the SM Higgs boson) is the only MSSM
Higgs boson that can be discovered at the LHC.  Further details can
be found in refs.~\cite{carenahaber,djouadi2}.

\section*{Acknowledgments}
\medskip

I would like to thank Jose Valle for providing me with the opportunity
to visit Valencia and to speak at the PASCOS-2010 conference.
I am especially grateful to Jose and his
colleagues at the Universitat de Val\`encia for their kind hospitality
during my visit.  I also appreciate a number of useful 
comments and suggestions from Sebastian Grab and
Lorenzo Ubaldi, who read the draft version of this manuscript.
This work is supported in part by the
U.S. Department of Energy, under grant number DE-FG02-04ER41268.

\end{document}